\newif\ifarxiv
\arxivtrue

\documentclass[12pt]{article}

\usepackage{graphicx,psfrag,amsmath,amsthm}
\usepackage{fullpage}
\newcommand{\BEAS}{\begin{eqnarray*}}
\newcommand{\EEAS}{\end{eqnarray*}}
\newcommand{\BEQ}{\begin{equation}}
\newcommand{\EEQ}{\end{equation}}
\newcommand{\BIT}{\begin{itemize}}
\newcommand{\EIT}{\end{itemize}}

\newcommand{\ie}{{\it i.e.}}

\newcommand{\ones}{\mathbf 1}
\newcommand{\reals}{{\mbox{\bf R}}}

\newcommand{\Tr}{\mathop{\bf Tr}}

\newcommand{\Co}{{\mathop {\bf Co}}} 

\newcommand{\argmin}{\mathop{\rm argmin}}

\newcounter{oursection}

\usepackage{array}
\usepackage{tikz}
\usepackage{multirow}
\usepackage{subcaption}

\title{A Distributed Method for Optimal Capacity Reservation}

\author{
Nicholas Moehle\thanks{Mechanical Engineering Department, Stanford University.
\texttt{moehle@stanford.edu}}
\and
Xinyue Shen\thanks{Electronic Engineering Department, Tsinghua University.}
\and 
Zhi-Quan Luo\thanks{Electrical and Computer Engineering, University of 
Minnesota.}
\and
Stephen Boyd\thanks{Electrical Engineering Department, Stanford University.}
}

\begin{document}

\maketitle

\begin{abstract}
We consider the problem of reserving link capacity in a network
in such a way that any of a given set of flow scenarios can be supported.
In the optimal capacity reservation problem, we choose the 
reserved link capacities to minimize the reservation cost.
This problem reduces to a large linear program, with the number of 
variables and constraints on the order of the number of links times 
the number of scenarios.
Small and medium size problems are within the capabilities of
generic linear program solvers.  
We develop a more scalable, distributed
algorithm for the problem that alternates between solving (in parallel)
one flow problem per scenario, and coordination steps, which 
connect the individual flows and the reservation capacities.
\end{abstract}

\section{Introduction}

We address the \emph{capacity reservation problem},
the problem of reserving link capacity in a network
in order to support multiple possible flow patterns.
We are given a description of the network,
and a collection of traffic demand scenarios,
which specify the amount of flow
to be routed from some sources to some sinks.
We must reserve enough link capacity so that for any of these scenarios,
there exist flows that route the traffic demand
while using no more than the reserved capacity on each link.
Subject to this requirement,
we seek to minimize a linear reservation cost.
This problem is also referred to as the \emph{network design problem}.

The capacity reservation problem can be expressed as a linear program (LP),
a fact first recognized by Gomory and Hu in 
1962 \cite{gomory1962application}.
For moderate problem sizes
it can be solved using generic LP solvers.
However, problems with a large number of scenarios
(and especially those that do not fit in a single computer's memory)
are beyond the reach of such solvers.

We present an iterative method for solving the capacity reservation problem
based on the alternating direction method of multipliers (ADMM).
Each iteration of the algorithm involves solving, in parallel,
a single minimum-cost flow problem for each scenario.
This means that our method can easily exploit parallel computing architectures,
and can scale to enormous problem instances
that do not fit in the memory of a single computer.
Unlike general ADMM,
our method maintains a feasible point at each iteration
(and therefore an upper bound on the problem value).
We can also compute a lower bound at modest cost,
which can be used to bound the suboptimality of the current iterate,
allowing for non-heuristic termination of the algorhtm.

\subsection{Previous work}
%

\paragraph{Capacity reservation problem with finite source set.}
The optimal reservation problem has a long history,
and was first proposed by Gomory and Hu in 1962
\cite{gomory1962application},
who consider the case of a finite source set,
and a single commodity.
In their extension to the multicommodity case,
they propose a decomposition method based on duality 
\cite{gomory1964synthesis}.
They also note that the structure of the capacity reservation problem makes
decomposition techniques attractive.
Labb\'{e} et al.\ \cite{labbe1999network}
propose two decomposition techinques,
one of which is based on Lagrangian duality
and foreshadows the method we propose here. 
Other decomposition approaches are based on cutting plane methods,
which can be interpreted as using only a small subset of the source set,
and adding additional source vectors as necessary.
Examples of this approach include
the work of Minoux et al.\ \cite{minoux1981optimum} and 
Petrou et al. \cite{petrou2007approach}.

\paragraph{Capacity reservation problem with infinite source set.}
Several other source set descriptions are given in the literature,
but are typically intractable.
For example, when the source set is given by polyhedron
(described by a set of inequalities),
the capacity reservation problem is known to be NP-hard 
\cite{chekuri2007hardness}.
However, for this and other cases,
such as ellipsoidal and conic source sets,
several effective heuristics exist.
One tractable approach is to restrict 
the flows to be an affine function of the demands.
This approach was first proposed by Ben-Ameur and Kerivin
\cite{ben2003networks,ben2005routing},
and is called 
\emph{static routing},
\emph{oblivious routing},
\emph{dynamic routing},
or \emph{affine routing},
depending on the form of the affine function used.
For a summary, see \cite{poss2011affine}.
Such a heuristic can be viewed as an application of 
affinely adjustable robust optimization,
has been studied extensively in robust optimization;
see \cite{ben2009robust}.


\paragraph{Convex optimization.}
In the case of a finite source set,
the capacity reservation problem can be reduced
to a convex optimization problem (indeed, a linear program).
These problems are tractable 
and mature software exists that can be used to solve them
\cite{boyd2004convex}.
Our approach to the capacity reservation problem
is based on a particular algortithm for solving convex optimization problems,
called the alternating direction method of multipliers (ADMM)
\cite{parikh2014proximal,boyd2011distributed},
which is well-suited for solving large convex problems
on distributed computers.


\subsection{Outline}
We start by describing the capacity reservation problem in \S\ref{s-cr},
and a simple heuristic for solving it.
In \S\ref{s-lower-bounds},
we provide a duality-based method for obtaining lower bounds
on the problem value.
In \S\ref{s-small-examples},
we illustrate the heuristic, as well as the lower bounds,
on two illustrative examples.
In \S\ref{s-admm}, we give an ADMM-based distributed algorithm for solving
the capacity reservation problem.
We conclude with a numerical example in \S\ref{s-num-example}.

\section{Capacity reservation problem}
\label{s-cr}

We consider a single-commodity flow on a network.
The network is modeled as a directed connected graph
with $n$ nodes and $m$ edges, described by its
incidence matrix $A\in \reals^{n\times m}$,
\[ 
A_{ij} = \left\{ \begin{array}{rl} 
1 & \mbox{edge $j$ points to node $i$}\\ 
-1 & \mbox{edge $j$ points from node $i$}\\ 
0 & \mbox{otherwise.} 
\end{array}\right. 
\] 
We let $f\in \reals^m$ denote the vector of edge flows,
which we assume are nonnegative and limited by a given edge capacity,
\ie, $0\leq f\leq c$, where the inequalities are interpreted 
elementwise, and $c\in \reals^m$ is the vector of edge capacities.
We let $s \in \reals^n$ denote the vector of sources, \ie,
flows that enter (or leave, when negative) the node.
Flow conservation at each node is expressed as $Af+s=0$.
(This implies that $\ones^Ts=0$.)
We say that $f$ is a \emph{feasible flow} for the source $s$
if there exists $f$ that satisfies $Af+s=0$ and $0\leq f \leq c$.
For a given source vector $s$,
finding a feasible flow (or determining that none exists)
reduces to solving a linear program (LP).

We consider a set of source vectors
$\mathcal S\subset \reals^n$, which we call the \emph{source set}.
A \emph{flow policy} is a function $\mathcal F:\mathcal S \to \reals^m$.
We say that a flow policy is feasible if for each $s\in \mathcal S$,
$\mathcal F(s)$ is a feasible flow for $s$.
Roughly speaking, a flow policy gives us a feasible flow for each
possible source.
Our focus is on the selection of a flow policy, given the 
network (\ie, $A$ and $c$)
and a description of the source set $\mathcal S$.

A \emph{reservation} (or reserved capacity)
is a vector $r\in\reals^m$ that satisfies $0 \leq r \leq c$.
We say that the reservation $r$ \emph{supports} a flow policy $\mathcal F$
if for every source vector $s\in \mathcal S$,
we have $\mathcal F(s) \leq r$.
Roughly speaking, we reserve a capacity $r_j$ on edge $j$; then,
for any possible source $s$, we can find a feasible flow $f$ that satisfies
$f_j \leq r_j$, \ie, it does not use more than the reserved capacity 
on each edge.
The cost of the reservation $r$ is
$p^Tr$, where $p\in \reals^m$, $p\geq 0$ is the vector of edge
reservation prices.
The \emph{optimal capacity reservation} (CR) problem is to find 
a flow policy $\mathcal F$ and a reservation $r$ that supports it, 
while minimizing the reservation cost.


The CR problem can be written as
\BEQ
\begin{array}{ll}
\mbox{minimize} & p^T r \\
\mbox{subject to} 
                 & A\mathcal F(s) + s = 0, \quad
                 0 \leq \mathcal F(s) \leq r,
                 \quad \forall s\in \mathcal S \\
                 & r \leq c.
\end{array}
\label{e-cr-prob}
\EEQ
The variables are the reservation $r\in\reals^m$
and the flow policy $F:\mathcal S \to \reals^m$.
The data are $A$, $\mathcal S$, $c$, and $p$.
This problem is convex, but when $\mathcal S$ is infinite,
it is infinite dimensional (since the variables include a function
on an infinite set), and contains a semi-infinite constraint,
\ie, linear inequalities indexed by an infinite set.
We let $J^\star$ denote the optimal value of the CR problem.

We will focus on the special case when $\mathcal S$ is finite,
$\mathcal S = \{s^{(1)},\dots,s^{(K)}\}$.
We can think of $s^{(k)}$ as the source vector in the $k$th
\emph{scenario}.
The CR problem (\ref{e-cr-prob}) is tractable in this case,
indeed, an LP:
\begin{equation}
\begin{array}{ll}
\mbox{minimize} & p^T r \\
\mbox{subject to} 
  & Af^{(k)} = s^{(k)}, \quad
 0 \leq f^{(k)} \leq r, \quad k = 1,\dots,K,\\
  & r \leq c.
\end{array}
\label{e-cr-lp}
\end{equation}
The variables are the reservations $r\in\reals^m$
and the scenario flow vectors $f^{(k)}\in\reals^m$
for each scenario $k$.
This LP has $m(K+1)$ scalar variables, $nK$ linear equality constraints,
and $m(2K+1)$ linear inequality constraints.
In the remainder of the paper,
we assume the problem data are such that this problem is feasible,
which occurs if and only if for each scenario there is a feasible
flow. (This can be determined by solving $K$ independent LPs.)

We note that a solution to (\ref{e-cr-lp}) is also a solution
for the problem (\ref{e-cr-prob}) with the 
(infinite) source set $\mathcal S = \Co \{s^{(1)},\ldots,s^{(K)}\}$,
where $\Co$ denotes the convex hull.
In other words, a reservation that is optimal for (\ref{e-cr-lp})
is also optimal for the CR problem in which the finite 
source set $\mathcal S$ is replaced with its convex hull.
To see this, we note that the optimal value of (\ref{e-cr-lp}) is 
a lower bound on the optimal value of (\ref{e-cr-prob}) with
$\mathcal S$ the convex hull, since the feasible set of the former 
includes the feasible set of the latter.
But we can extend a solution of the LP (\ref{e-cr-lp}) to a feasible
point for the CR problem (\ref{e-cr-prob}) with the same objective,
which therefore is optimal.
To create such as extension, we define $\mathcal F(s)$ for any 
$s \in \mathcal S$.
We define $\theta(s)$ as the unique least Euclidean norm 
vector with $\theta(s) \geq 0$, $\ones^T\theta(s) =1$ (with $\ones$ the vector 
with all entries one) that satisfies $s = \sum_k \theta_k s^{(k)}$.
(These are barycentric coordinates.)
We then define
$\mathcal F(s) = \sum_{k=1}^K \theta_k(s) f^{(k)}$.  

We note for future use another form of the CR problem~(\ref{e-cr-lp}).
We eliminate $r$ using $r = \max_k f^{(k)}$, where the max is understood to 
apply elementwise.
Then we obtain the problem
\begin{equation}
\begin{array}{ll}
\mbox{minimize} & p^T \max_k f^{(k)} \\
\mbox{subject to} 
  & Af^{(k)} = s^{(k)}, \quad
 0 \leq f^{(k)} \leq c, \quad k = 1,\ldots,K,
\end{array}
\label{e-cr-2}
\end{equation}
where the variables are $f^{(k)}$, $k=1, \ldots, K$.
This is a convex optimization problem.

We will also express the CR problem using matrices, as
\begin{equation}
\begin{array}{ll}
\mbox{minimize} & p^T \max(F) \\
\mbox{subject to} & AF + S = 0, \\
                  & 0 \leq F \leq C
\end{array}
\label{e-cr-mat}
\end{equation}
with variable $F\in\reals^{m\times K}$.
The columns of $F$ are the flow vectors $f^{(1)},\ldots,f^{(K)}$.
The columns of $S$ are the source vectors
$s^{(1)}, \ldots, s^{(K)}$,
and the columns of $C$ are all the capacity vector $c$, \ie,
$C = c\ones^T$, where $\ones$ is the vector with all entries one.
The inequalities above are interpreted elementwise, and
the $\max$ of a matrix
is taken over the rows,
\ie, it is the vector containing the largest element in each row.

\paragraph{Heuristic flow policy.}
\label{s-heuristic}
We mention a simple heuristic for the
reservation problem that foreshadows our method.
We choose each scenario flow vector $f^{(k)}$
as a solution to the capacitated minimum-cost flow problem
\begin{equation}
\begin{array}{ll}
\mbox{minimize} & p^T f \\
\mbox{subject to} 
                  & Af = s^{(k)}, \quad
                    0 \leq f \leq c,
\end{array}
\label{e-naive-prob}
\end{equation}
with variable $f\in\reals^m$.
These flow vectors are evidently feasible
for (\ref{e-cr-2}),
and therefore the associated objective value $J^\mathrm{heur}$
is an upper bound on the optimal value of the problem,
\ie,
\[
J^\star \leq J^\mathrm{heur} = p^T \max_k f^{(k)}.
\]
Note that finding the flow vectors using this heuristic
involves solving $K$ smaller LPs independently,
instead of the one large LP (\ref{e-cr-lp}).
This heuristic flow policy greedily minimizes the cost
for each source separately; but it does not coordinate the 
flows for the different sources to reduce the reservation cost.


We will now show that we also have
\[
J^\mathrm{heur}/K \leq J^\star,
\]
\ie,
$J^\mathrm{heur}/K$ is a lower bound on $J^\star$.  This shows that
the heuristic policy is no more than $K$-suboptimal.
We first observe that for any nonnegative vectors $f^{(k)}$, we have
\BEQ
\label{e-heur-lb}
(1/K) \sum_{k=1}^K p^T f^{(k)}
=
p^T \left((1/K)\sum_{k=1}^K f^{(k)}\right)
\leq
p^T \max_{k} f^{(k)},
\EEQ
which follows from the fact that the maximum of a set of $K$ 
numbers is at least as big as their mean.  
Now minimize the lefthand and righthand sides over
the feasible set of (\ref{e-cr-lp}).
The the righthand side becomes $J^\star$,
and the lefthand side becomes $J^\mathrm{heur}$.

In \S\ref{ex-k-subopt} we give an example showing
that this suboptimality bound is tight,
\ie, the heuristic flow policy can indeed be up to $K$ suboptimal.

\section{Lower bounds and optimality conditions}
\label{s-lower-bounds}
The heuristic flow policy described above provides an upper and lower bound
on $J^\star$.
These bounds are computationally appealing because they involve
solving $K$ small LPs independently.
Here we extend the basic heuristic method to one in which
the different flows are computed using different prices.
This gives a parametrized family of lower and upper
bounds on $J^\star$.
Using ideas based on linear programming duality, we can show that
this family is tight, \ie, there is a choice of 
prices for each scenario that yields lower and upper bound $J^\star$.
This duality idea will form the basis
for our distributed solution to (\ref{e-cr-lp}),
described in \S\ref{s-admm}.

\paragraph{Scenario prices.}
We consider nonnegative \emph{scenario pricing vectors}
$\pi^{(k)}\in \reals^m$, for $k=1,\ldots, K$,
satisfying
$\sum_{k=1}^K \pi^{(k)} = p$.

Then we have
\begin{equation}
p^T \max_k f^{(k)} \geq \sum_{k=1}^K \pi^{(k)T} f^{(k)}.
\label{e-inequality}
\end{equation}
It follows that the optimal value of the LP
\BEQ
\begin{array}{ll}
\mbox{minimize} & \sum_{k=1}^K \pi^{(k)T} f^{(k)}  \\
\mbox{subject to} 
                  & Af^{(k)} = s^{(k)}, \quad
                    0 \leq f^{(k)} \leq c, \quad k = 1,\ldots,K
\end{array}
\label{e-cr-relax}
\EEQ
is a lower bound on $J^\star$.
This problem is separable;
it can be solved by solving $K$ small LPs (capacited flow problems)
independently.
In other words, to obtain the bound,
we first decompose the reservation prices on each edge into scenario prices.
Then, for each scenario $k$,
the flow $f^{(k)}$ is chosen 
(independently of the other scenarios)
to be a minimum-cost flow
according to price vector $\pi^{(k)}$.
The scenario price vectors can be interpreted
as a decomposition of the reservation price vector,
\ie, the price of reservation along each edge can be decomposed
into a collection of flow prices for that edge,
with one for each scenario.

Note that the inequality 
(\ref{e-heur-lb})
is a special case of 
(\ref{e-inequality}),
obtained with scenario prices $\pi^{(k)} = (1/K)p$.
In fact, with these scenario prices,
the heuristic subproblem (\ref{e-naive-prob})
is also just a special case of (\ref{e-cr-relax}),
\ie, the family of lower bounds obtained by solving (\ref{e-cr-relax})
with different scenario pricing vectors
includes the single lower bound obtained
using the heuristic method.

Given a set of scenario prices, the flows found as above
are feasible for the CR problem.  Therefore $p^T \max_k f^{(k)}$
is an upper bound on $J^\star$.

\paragraph{Tightness of the bound and optimality conditions.}
\label{s-opt-cond}
There exist some scenario prices
for which the optimal value $J^\star$ of the CR problem (\ref{e-cr-lp})
and its relaxation (\ref{e-cr-relax}) are the same.
This is a consequence of linear programming duality;
a (partial) dual of (\ref{e-cr-lp}) is the problem
of determining the scenario pricing vectors that maximize
the optimal value of (\ref{e-cr-relax}).
In addition,
given such scenario prices,
any optimal flow policy for (\ref{e-cr-lp})
is also optimal for (\ref{e-cr-relax})
with these scenario prices.

We can formalize this notion as follows.
A flow policy $f^{(k)}$
is optimal if and only if
there exists a collection of scenario prices $\pi^{(k)}$
such that the following conditions hold.
\begin{enumerate}
\item
\label{c-price-decomp}
The vectors $\pi^{(k)}$ must be valid scenario prices, \ie,
\begin{equation}
\sum_{k=1}^K \pi^{(k)}_j = \pi_j,
\quad
\pi^{(k)} \geq 0,
\quad k=1,\ldots,K.
\label{e-opt-prices}
\end{equation}
\item
\label{c-zero-gap}
The inequality (\ref{e-inequality}) must be tight:
\[
p^T \max_k f^{(k)} = \sum_{k=1}^K \pi^{(k)T} f^{(k)}.
\]
\item 
\label{c-min-cost-flow}
The flow vectors $f^{(k)}$, for $k=1,\ldots,K$,
must be optimal for 
(\ref{e-naive-prob}),
with scenario prices $\pi^{(k)}$.
\end{enumerate}



Condition~\ref{c-zero-gap} has the interpretation that
the optimal reservation cost
can be written as a sum of $K$ cost terms
corresponding to the scenarios,
\ie, the reservation cost can be allocated to the $K$ scenarios.
This can be made the basis for a payment scheme,
in which scenario $k$ is charged with a fraction 
$\pi^{(k)T} f^{(k)}$
of the total reservation cost.
Note that because the scenario prices 
that show optimality of a collection of scenario flow vectors
are not necessarily unique,
these cost allocations may not be unique.

Condition~\ref{c-zero-gap} also implies the complimentarity condition
\[
\pi_j^{(k)T} (f^{(k)} - r)_j = 0
\]
for all $k$ and $j$.
In other words,
a positive scenario price on an edge implies
that the corresponding scenario edge flow is equal to the optimal reservation,
\ie, this scenario edge flow must contribute to the reservation cost.
Similarly,
if a scenario edge flow is not equal to that edge's reservation,
then the scenario price must be zero for that edge.

Condition~\ref{c-min-cost-flow} means that,
given the scenario prices,
the flow vector for each scenario is a minimum-cost flow
using the scenario price vector associated with that scenario.
We can interpret the price vectors as weighting the contribution
of a scenario edge flow to the total reservation cost.
In the case that $\pi_j^{(k)}=0$,
then the flow over edge $j$ under scenario $k$ does not contribute
to the total reservation cost.
If $\pi_j^{(k)}>0$,
then the flow over edge $j$ under scenario $k$ contributes
to the total reservation cost,
and from before, we have $f^{(k)}_j = r_j.$
Additionally, if $\pi^{(k)} = 0$ for some $k$,
then scenario $k$ is irrelevant
\ie, removing it has no effect on the reservation cost.
If instead $\pi^{(k)} = p$,
we conclude that scenario $k$ is the only relevant scenario.

\section{Examples}
\label{s-small-examples}

\subsection{Illustrative example}
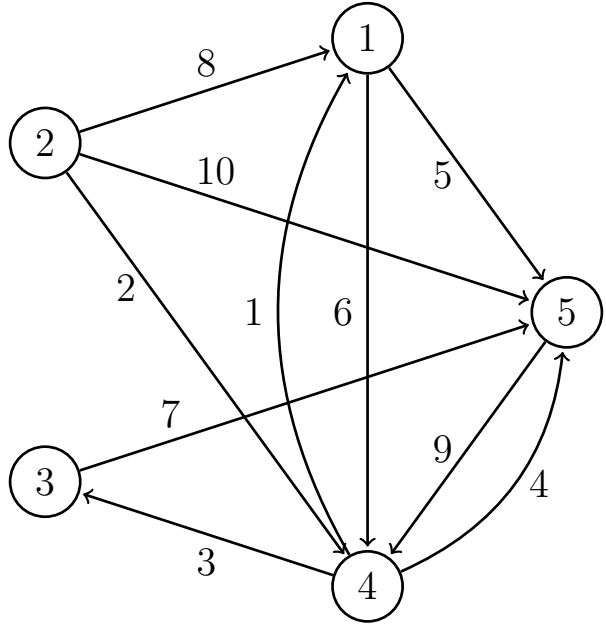
\begin{figure}
\begin{center}
\resizebox{.5\textwidth}{!}{
  \begin{tikzpicture}[->,shorten >=1pt,auto,node distance=3cm,thick,
                  main node/.style={circle,draw},
                  sourcenode/.style={},
                  sourceedge/.style={dashed}]

    \def\DIR{72,144,216,288,360}
    \foreach \d [count=\i] in \DIR{
      \node[main node] at (\d:3) (\i) {\i};
    }

    \path (4) edge[pos=.50, left ,bend left ] node { 1} (1);
    \path (2) edge[pos=.30, left ,          ] node { 2} (4);
    \path (4) edge[pos=.50, below,          ] node { 3} (3);
    \path (4) edge[pos=.50, right,bend right] node { 4} (5);
    \path (1) edge[pos=.50, left ,          ] node { 5} (5);
    \path (1) edge[pos=.50, left ,          ] node { 6} (4);
    \path (3) edge[pos=.20, above,          ] node { 7} (5);
    \path (2) edge[pos=.50, above,          ] node { 8} (1);
    \path (5) edge[pos=.50, left ,          ] node { 9} (4);
    \path (2) edge[pos=.30, above,          ] node {10} (5);




  \end{tikzpicture}
}
\end{center}
\caption{The graph of the illustrative example.}
\label{f-ex1-graph}
\end{figure}

Here we consider a simple example with
$n = 5$ nodes,
$m = 10$ edges,
and $K = 8$ scenarios.
The graph was generated randomly, from a uniform distribution
on all connected graphs with $5$ nodes and $10$ edges,
and is shown in figure~\ref{f-ex1-graph}.
We use price vector $p = \ones$ and capacity vector $c = \ones$.
The $8$ scenario source vectors were chosen according to
$s^{(k)} = -Az^{(k)}$,
where each element of $z^{(k)}$ was randomly chosen from
$\{0,1/3,2/3,1\}$.
This guarantees that there is a feasible flow for each scenario.

\begin{figure}
\begin{center}
\ifarxiv
  \includegraphics[width=.9\textwidth]{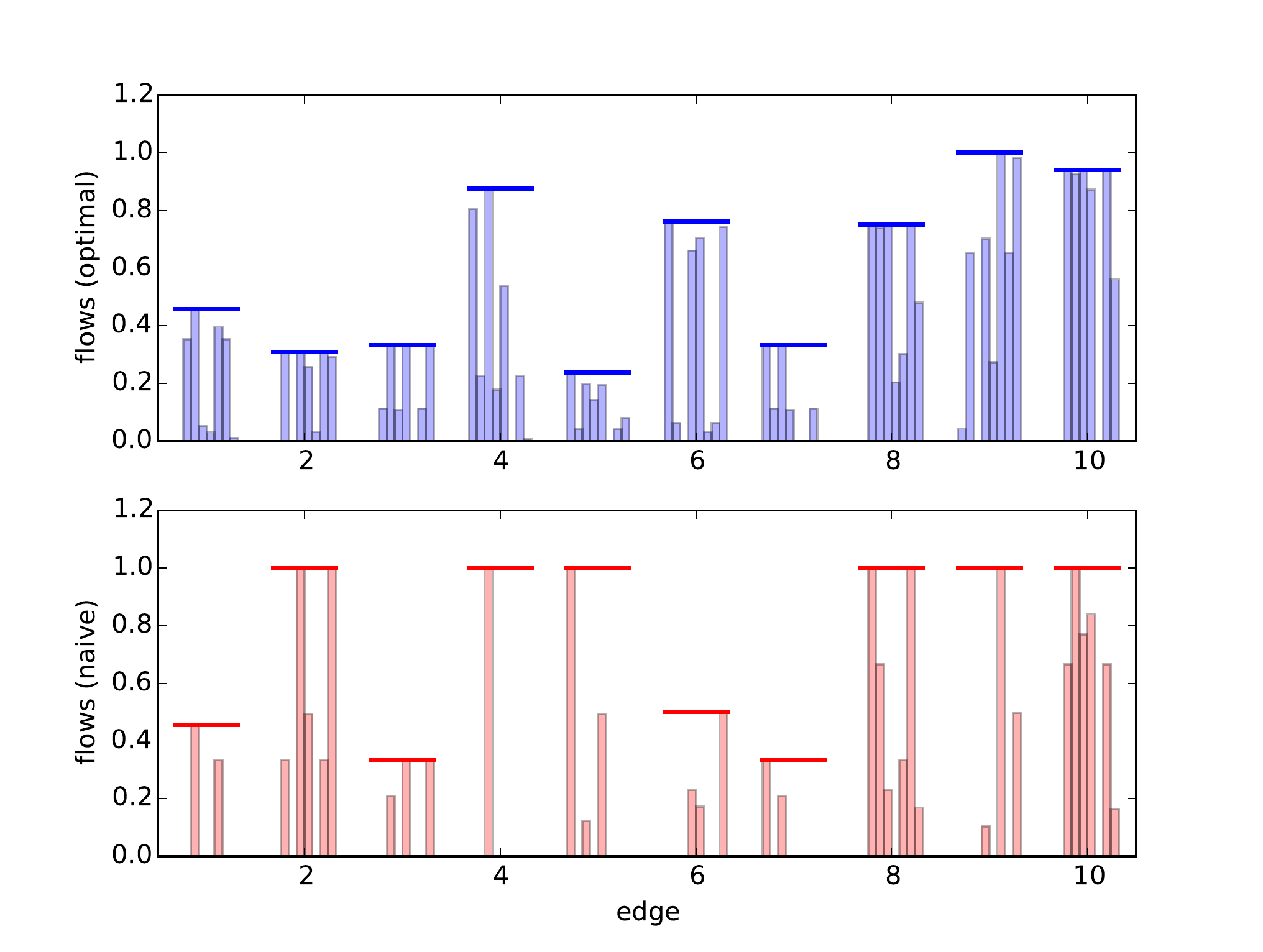}
\else
  \includegraphics[width=.9\textwidth]{figures/edge_flows.pdf}
\fi
\caption{The optimal and naive flow policies.}
\label{f-edge-flows}
\end{center}
\end{figure}

The optimal reservation cost is $6.0$,
and the objective of the heuristic policy is $7.6$.
(The lower bound from the heuristic policy is $2.3$.)
The optimal and heuristic flow policies
are shown in figure (\ref{f-edge-flows}).
The upper plot shows the optimal policy,
and the lower plot shows the heuristic policy.
For each plot, the bars show the flow policy;
the $10$ groups are the edges, and the $8$ bars
are the edge flows under each scenario.
The line above each group of bars is the reservation for that edge.

Optimal scenario prices are given in table~\ref{t-ex1-prices}.
First, we confirm that the row sums are indeed $p$.
The fact that only some scenario prices are positive for each edge
indicates that no scenario is relevant for every edge,
\ie, that there is no single scenario that is being planned for.
However,
for every scenario, 
the scenario prices are positive for at least one edge,
which means that every scenario is potentially relevant to the optimal reservation.

\begin{table}
	\begin{center}
		\begin{tabular}{cc|cccccccc}
			     & & \multicolumn{8}{c}{Scenario}    \\
			     & & 1 & 2 & 3 & 4 & 5 & 6 & 7 & 8  \\
			\hline 
      \multirow{10}{*}{\rotatebox[origin=c]{90}{Edge}}
      & 1  &      &      & 1.0  &      &      &     &      &      \\
      & 2  &      & 0.33 &      & 0.33 &      &     & 0.33 &      \\
      & 3  &      &      & 0.38 &      & 0.28 &     &      & 0.33 \\
      & 4  &      &      & 1.0  &      &      &     &      &      \\
      & 5  & 1.0  &      &      &      &      &            &      \\
      & 6  & 1.0  &      &      &      &      &     &      &      \\
      & 7  & 0.38 &      & 0.62 &      &      &     &      &      \\
      & 8  &      & 0.33 &      & 0.33 &      &     & 0.33 &      \\
      & 9  &      &      &      &      &      & 1.0 &      &      \\
      & 10 &      & 0.33 &      & 0.33 &      &     & 0.33 &      \\
		\end{tabular}
	\end{center}
	\caption{
  The scenario prices $\pi^{(1)},\dots,\pi^{(8)}$.
  Column $k$ in the table is the scenario price vector $\pi^{(k)}$.
  Blank spaces correspond to $0$ entries.
  }
	\label{t-ex1-prices}
\end{table}

\subsection{$K$-suboptimal example}
\label{ex-k-subopt}
In \S\ref{s-heuristic},
we showed that the heuristic method
is not more than $K$ suboptimal,
and the previous example showed that the heuristic approach is
not, in fact, optimal.
Here we give an example that shows that the $K$-suboptimality
result is in fact tight,
\ie, for every positive integer $K$,
there is an instance of (\ref{e-cr-lp}) for which the heuristic
is (nearly) $K$ suboptimal.

\begin{figure}
\begin{center}
\resizebox{.6\textwidth}{!}{
  \begin{tikzpicture}[->,shorten >=1pt,auto,node distance=2cm,thick,
                  main node/.style={circle,draw},
                  sourcenode/.style={},
                  sourceedge/.style={dashed}]

    \node[main node] at (0,0)  (1)               {1};
    \node[main node]           (2)  [below of=1] {2};
    \node[main node]           (3)  [below of=2] {3};

    \node[main node] at (5,0)  (4)               {4};
    \node[main node]           (5)  [below of=4] {5};
    \node[main node]           (6)  [below of=5] {6};

    \node[main node] at (10,-2) (7)              {7};

    \path
      (1) edge node [pos=.40, above]{ $\epsilon$} (4)
      (1) edge node [pos=.30, above]{$2\epsilon$} (5)
      (1) edge node [pos=.05, below]{$2\epsilon$} (6);
    \path
      (2) edge node [pos=.10, above]{$2\epsilon$} (4)
      (2) edge node [pos=.30, above]{ $\epsilon$} (5)
      (2) edge node [pos=.10, below]{$2\epsilon$} (6);
    \path
      (3) edge node [pos=.05, above]{$2\epsilon$} (4)
      (3) edge node [pos=.30, below]{$2\epsilon$} (5)
      (3) edge node [pos=.40, below]{ $\epsilon$} (6);
    \path
      (4) edge node [above]{$1$} (7)
      (5) edge node [above]{$1$} (7)
      (6) edge node [above]{$1$} (7);
  \end{tikzpicture}
}
\end{center}
\caption{
Graph example for $a=3$.
The labels give the reservation price for that edge.
}
\label{f-ex2-graph}
\end{figure}
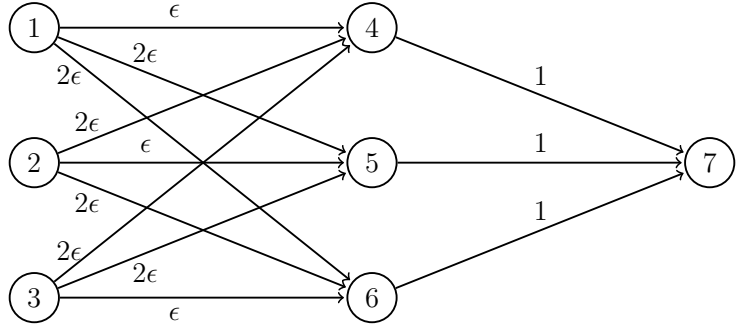

We consider a graph with $n = 2a+1$ nodes,
for some positive integer $a$,
with the nodes arranged in three layers,
with $a$, $a$, and $1$ nodes, respectively.
(We show the graph for $a=3$ in figure~\ref{f-ex2-graph}.)
There is an edge connecting 
every node in the first layer
with every node in the second layer,
\ie, for every $i = 1,\dots,a$
and every $j = a+1,\dots, 2a$,
the edge $(i,j)$ is in the graph.
The price in this edge is $\epsilon$ if $i=j$,
and $2\epsilon$ if $i\neq j$.
Similarly, every node in the second layer connects to the last node,
\ie, for every $i=a+1,\dots,2a$,
the edge $(i,2a+1)$ is in the graph.
The price for this edge is $1$.

We consider $K = a$ scenarios.
In each scenario,
one unit of flow in injected into a single
node in the first layer,
and must be routed to the last node.
In other words, the source vector $s^{(k)}$ is all zeros,
except for the $k$th element, which is $1$,
and the last element, which is $-1$.

A feasible (but not necessarily optimal) 
point for the CR problem (\ref{e-cr-lp}) is
to route all flow equally through node $a+1$,
the first node in the second layer.
We must then reserve one unit of capacity for edge $(a+1, 2a+1)$,
and one unit of capacity from each node in the first layer
to node $a+1$,
\ie, for edges $(i,a+1)$, for $i=1,\dots, a$.
The total reservation cost is $(2a - 1)\epsilon +1$.

We now consider the (unique) heuristic solution.
The solution to (\ref{e-naive-prob}) under scenario $k$ is to route all flow
from node $k$ to node $a+k$, and finally to node $2a+1$,
Because no edges support flow under any two different scenarios,
the heuristic therefore achieves a reservation cost of $K(\epsilon +1)$.

As $\epsilon \to 0$,
the cost difference between the feasible reservation for (\ref{e-cr-lp})
and the heuristic reservation approaches $K$.
Because the feasible reservation provides
an upper bound on the optimal reservation cost,
the difference between the optimal
reservation and the heuristic reservation
also approaches $K$.

\section{Distributed solution method}
\label{s-admm}
In this section we provide a parallelizable algorithm,
based on the alternating direction method of multipliers (ADMM),
to solve \eqref{e-cr-lp}.
(For details on ADMM, see \cite{boyd2011distributed}.)
We first express the problem~(\ref{e-cr-mat}) in the 
consensus form
\begin{equation}
\begin{array}{ll}
\mbox{minimize} & p^T \max(\tilde F) + g(F) \\
\mbox{subject to} & F = \tilde F
\end{array}
\label{e-consensus-prob}
\end{equation}
with variables $F\in\reals^{m\times K}$ and
$\tilde F\in\reals^{m\times K}$.
The function $g$ is the indicator function for feasible
flows,
\[
g(F) =
\left\{ \begin{array}{ll} 
0 & AF + S = 0, \quad 0 \leq F \leq C \\ 
\infty & \mbox{otherwise.}
\end{array}\right. 
\]
In the consensus problem (\ref{e-consensus-prob})
we have replicated the flow matrix variable,
and added a consensus constraint, \ie, the requirement
that the two variables representing flow policy must agree.

The augmented Lagrangian for problem (\ref{e-consensus-prob}) is
\[
L(F,\tilde F, \Pi) = p^T \max (\tilde F) + g(F)
+ \Tr \Pi^T (F - \tilde F) + (\rho/2) \|F - \tilde F \|_F^2,
\]
where $\|\cdot\|_F$ is the Frobenius norm,
and $\rho > 0$ is a parameter.
Here $\Pi\in\reals^{m\times K}$ is the dual variable.


\paragraph{ADMM.}
The ADMM algorithm (with over-relaxation) for (\ref{e-consensus-prob})
is given below.
First we initialize the iterates $\tilde F(0)$ and $\Pi(0)$.
Then we carry out the following steps:
\begin{align*}
F(l+1) &= \argmin_{F} L\big(F, \tilde F(l), \Pi(l)\big) \\
\tilde F(l+1) &= \argmin_{\tilde F} 
L\big(\alpha F(l+1) + (1-\alpha) \tilde F(l), \tilde F, \Pi(l)\big) \\
\Pi(l+1) &= \Pi(l) + 
\rho \big(\alpha F(l+1) + (1-\alpha) \tilde F(l) - \tilde F(l+1)\big),
\end{align*}
where $\alpha$ is an algorithm parameter in $(0,2)$;
the argument $l$ is the iteration number.
For reasons that will become clear,
we call these three steps the \emph{flow policy update},
the \emph{reservation update},
and the \emph{price update}, respectively.

\paragraph{Convergence.}
Given our assumption that (\ref{e-cr-lp}) is feasible,
we have
$F(l)\to F^\star$, $\tilde F(l)\to F^\star$,
and
$\Pi(l) \to \Pi^\star$,
where $(F^\star,\Pi^\star)$ is a primal-dual solution to (\ref{e-cr-mat}).
This follows from standard results on ADMM 
\cite{boyd2011distributed}.

\paragraph{Flow policy update.}
Minimizing the Lagrangian over $F$
requires minimizing a quadratic function
over the feasible set of (\ref{e-cr-mat}).
This can be interpreted as the solution
of a minimum-cost flow problem for each of the $K$ scenarios.
More specifically,
column $k$ of the matrix $F(l+1)$ is the unique solution
to the quadratic minimum-cost flow problem:
\begin{equation}
\begin{array}{ll}
\mbox{minimize} & \pi^{(k)T} f + (\rho/2)\|f - \tilde{f}^{(k)}\|^2 \\
\mbox{subject to} 
                  & Af + s^{(k)} = 0 \\
                  & 0 \leq f \leq c. 
\end{array}
\label{e-flow-subprob}
\end{equation}
Here $\tilde f^{(k)}$ and $\pi^{(k)}$ are the $k$th columns
of the matrices $\tilde F(l)$, and $\Pi(l)$, respectively.
These $K$ flow problems can be solved in parallel
to update the $K$ columns of the iterate $F(l+1)$.
Note that the objective can be interpreted
as a sum of the edge costs,
using the estimate of the scenario prices,
plus a quadratic regularization term that penalizes deviation
of the flow vector from the previous flow values $\tilde f^{(k)}$.

\paragraph{Reservation update.}
Minimizing the Lagrangian over $\tilde F$
requires minimizing the sum of a quadratic function of $\tilde F$
plus a positive weighted sum of its row-wise maxima.
This step decouples into $m$ parallel steps.
More specifically,
The $j$th row of $\tilde F(l+1)$ is the unique solution to
\begin{equation}
\begin{array}{ll}
\mbox{minimize} & p_j \max \big(\tilde{f}\big)
- \pi_j^T \tilde{f} 
+ (\rho/2) \left\|\tilde{f} - \big(\alpha f_j + (1-\alpha) \tilde f_j\big) \right\|^2, \\
\end{array}
\label{e-res-update}
\end{equation}
where $\tilde f$ is the variable, and
$f_j$, $\tilde f_j$, and $\pi_j$ are the $j$th rows of $F(l+1)$, $\tilde F(l)$, 
and $\Pi(l)$, respectively.
(The scalar $p_j$ is the $j$th element of the vector $p$.)
We interpret the $\max$ of a vector to be the largest element of that vector.
This step can be interpreted as an implicit update of the reservation vector.
The first two terms are the difference
between the two sides of (\ref{e-inequality}).
The problem can therefore be interpreted as finding flows for each edge
that minimize the looseness of this bound
while not deviating much from a combination of
the previous flow values $f_j$ and $\tilde f_j$.
(Recall that at a primal-dual solution to (\ref{e-cr-lp}),
this inequality becomes tight.)
There is a simple solution for each of these subproblems,
given in appendix~\ref{a-res-update};
the computational complexity for each subproblem scales like $K\log K$.

\paragraph{Price update.}
In the third step, the dual variable $\Pi(l+1)$ is updated.

\paragraph{Iterate properties.}
Note that the iterates $F(l)$, for $l = 1,2\dots$,
are feasible for (\ref{e-cr-mat}),
\ie, the columns of $F(l)$ are always feasible flow vectors.
This is a result of the simple fact that each column
is the solution of the quadratic minimum-cost flow problem (\ref{e-flow-subprob}).
This means that $U(l) = p^T\max F(l)$ is an upper bound
on the optimal problem value.
Furthermore, because $F(l)$ converges to a solution of (\ref{e-cr-mat}),
this upper bound $U(l)$ converges to $J^\star$
(though not necessarily monotonically).

It can be shown that $\Pi(l)\geq 0$  and $\Pi(l)\ones = p$
for $l=1,2,\dots$,
\ie, the columns of $\Pi(l)$ are feasible scenario price vectors
satisfying optimality condition \ref{c-price-decomp} of 
\S\ref{s-opt-cond}.
This is proven in Appendix~\ref{a-iterate-prop}.
We can use this fact to obtain a lower bound on the problem value,
by computing the optimal value of (\ref{e-cr-relax}),
where the scenario pricing vectors $\pi^{(k)}$ are given
by the columns of $\Pi(l)$,
which is a lower bound on the optimal value of (\ref{e-cr-mat}).
We call this bound $L(l)$.
Because $\Pi(l)$ converges
to an optimal dual variable of (\ref{e-consensus-prob}),
$L(l)$ converges to $J^\star$
(though it need not converge monotonically).

Additionally, the optimality condition \ref{c-zero-gap}
is satisfied by iterates $\tilde F(l+1)$ and $\Pi(l+1)$,
for $l=1,2,\dots$,
if we take $f^{(k)}$ and $\pi^{(k)}$ to be the columns
$\tilde F(l+1)$ and $\Pi(l+1)$, respectively.
This is shown in Appendix~\ref{a-iterate-prop}.


\paragraph{Stopping criterion.}
Several reasonable stopping criteria are known for ADMM.
One standard stopping criterion is that
that the norms of the primal residual $\|F(l) - \tilde F(l)\|$
and the dual residual $\|\Pi(l+1) - \tilde \Pi(l)\|$ are small.
(For a discussion, see
\cite[\S3.3.1]{boyd2011distributed}.)

In our case,
we can use the upper and lower bounds $U(l)$ and $L(l)$
to bound the suboptimality of the current feasible point $F(l)$.
More specifically,
we stop when 
\begin{equation}
U(l)-L(l) \leq \epsilon^\mathrm{rel} L(l)
\label{e-stopping-crit}
\end{equation}
which guarantees a relative error not exceeding $\epsilon^\mathrm{rel}$.
Note that computing $L(l)$ requires solving $K$ small LPs,
which has roughly the same computational cost as one ADMM iteration;
it may therefore be beneficial to check this stopping criterion
only occasionally, instead of after every iteration.

\paragraph{Parameter choice.}
Although our algorithm converges for any positive $\rho$
and any $\alpha\in(0,2)$,
the values of these parameters strongly impact the number
of iterations required.
In many numerical experiments, we have found that
choosing $\alpha = 1.8$ works well.
The choice of the parameter $\rho$ is more critical.
Our numerical experiments suggest choosing $\rho$ as
\begin{equation}
\rho = \mu \frac{ \ones^T p}{\max(\ones^T F^\mathrm{heur})},
\label{e-rho-selection}
\end{equation}
where $F^\mathrm{heur}$ is a solution to the heuristic problem
described in \S\ref{s-heuristic},
and $\mu$ is a positive constant.
With this choice of $\rho$,
the primal iterates $F(l)$ and $\tilde F(l)$
are invariant under positive scaling of $p$,
and scale with $S$ and $c$.
(Similarly, the dual iterates $\Pi(l)$ scale with $p$,
but are invariant to scaling of $S$ and $c$.)
Thus the choice~(\ref{e-rho-selection}) renders the 
algorithm invariant to scaling of $p$, and also to 
scaling of $S$ and $c$.
Numerical experiments suggest that values of 
$\mu$ in the range between $0.01$ and $0.1$
work well in practice.

\paragraph{Initialization.}
Convergence of $F(l)$, $\tilde F(l)$, and $\Pi(l)$
to optimal values is guaranteed for any initial choice
of $\tilde F(0)$ and $\Pi(0)$.
However we propose initializing
$\tilde F(0)$ as a solution to the heuristic method described in 
\S\ref{s-heuristic}, and $\Pi(0)$ as $(1/K)p\ones^T$.
(In this case, we have $F(1) = \tilde F(0)$,
which means the initial feasible point is no more than $K$ suboptimal.)


\section{Numerical example}
\label{s-num-example}

\paragraph{Problem instance.}
We consider a numerical example with $n = 2000$ nodes, $m = 5000$ edges, 
and $K = 1000$ scenarios,
for which the total number of (flow) variables is 5 million.
The graph is randomly chosen from a uniform distribution on
all graphs with $2000$ nodes and $5000$ edges.
The elements of the price vector $p$ are uniformly randomly distributed
over the interval $[0,1]$, and $c = \ones$.
The elements of the source vectors are generated according to 
$s^{(k)} = -A z^{(k)}$, where the elements of $z^{(k)}$ are uniformly randomly 
distributed over the interval $[0,1]$.
The optimal value is $18053$,
and the objective obtained by the heuristic flow policy is $24042$.
The initial lower bound is $5635$,
so the initial relative gap is $(U(0)-L(0))/L(0)=3.26$.

\paragraph{Convergence.}
We use $\alpha = 1.8$, with
$\rho$ chosen according to (\ref{e-rho-selection})
with $\mu = 0.05$.
We terminate when (\ref{e-stopping-crit}) is satisfied
with $\epsilon^\mathrm{rel} = 0.01$, \ie, when we can guarantee
that the flow policy is no more than $1\%$ suboptimal.
We check the stopping criterion at every iteration;
however, we only update the lower bound (\ie, evaluate $L(l)$)
when $l$ is a multiple of $10$.

\begin{figure}
\centering
\ifarxiv
  \includegraphics[width=0.9\textwidth]{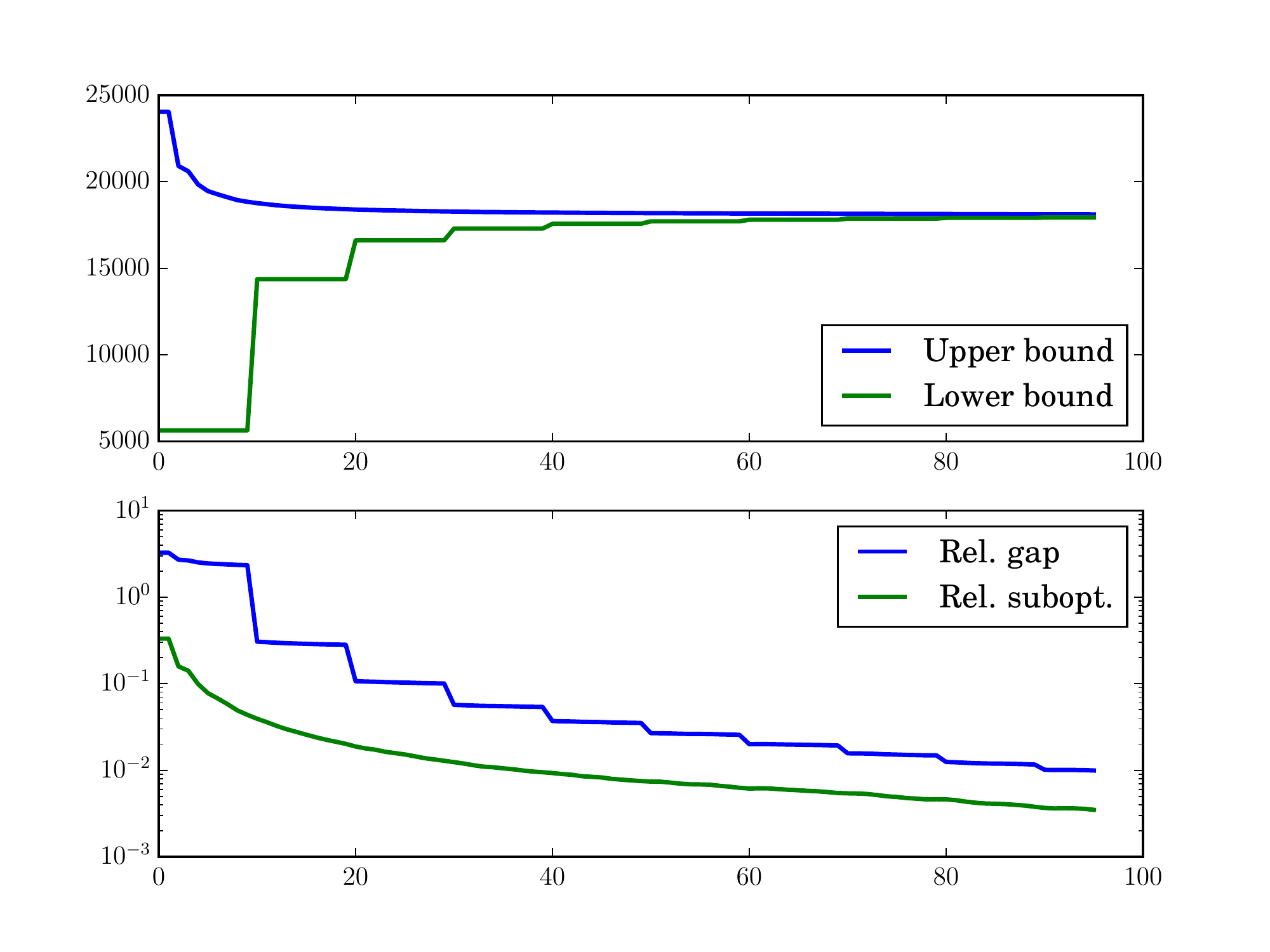}
\else
  \includegraphics[width=0.9\textwidth]{figures/bounds.pdf}
\fi
\caption{
   \emph{Top}: Upper and lower bounds $U(l)$ and $L(l)$ on the optimal value.
 \emph{Bottom}: Estimated relative gap $(U(l) - L(l))/L(l)$ 
 and relative suboptimality $(U(l) - J^\star)/J^\star$.}
\label{f-bounds}
\end{figure}

The algorithm terminates in $95$ iterations.
The convergence is shown in figure~\ref{f-bounds}.
The upper plot shows the convergence of the upper and lower bounds
to the optimal problem value,
and the lower plot shows the relative gap $(U(l) - L(l))/L(l)$
and the realtive suboptimality
$(U(l) - J^\star)/J^\star$ (computed after the algorithm 
has converged to very high accuracy).
(Note that both the upper and the lower bounds 
were updated by the best values to date.
\ie, they were non-increasing and non-decreasing, 
respectively.
The lower bound was only updated once every $10$ iterations.)
Our algorithm guarantees that we terminate with a policy
no more than $1\%$ suboptimal; in fact,
the final policy is around $0.4\%$ suboptimal.

\paragraph{Timing results.}
We implemented our method in Python,
using the multiprocessing package.
The flow update subproblems were solved using
the commercial interior-point solver Mosek \cite{mosek},
interfaced using CVXPY,
a Python-based modeling language for convex optimization \cite{cvxpy}.
The reservation update subproblems were solved using the simple method
given in Appendix~\ref{a-res-update}.
(The simulation was also run using the open source solver
ECOS \cite{domahidi2013ecos}, which however is single-threaded, and unable to solve
the full problem directly.  Mosek is able to solve the full problem
directly, which allows us to compare the solve time with our
distributed method.)

We used a Linux computer with $4$ Intel Xeon E5-4620 CPUs,
with $8$ cores per CPU and $2$ virtual cores per physical core.
The flow update and reservation update subproblems
were solved in parallel with $64$ hyperthreads.

The computing times are listed in table \ref{optV-runtime}. 
The first line gives the average time to solve a single
flow update subproblem (\ref{e-flow-subprob}).
The second line gives the total time taken to solve the problem using
our distributed method (which took $95$ iterations).
The third line gives the time taken to solve the full CR problem
directly.
For this simulation, we terminated Mosek when the 
relative surrogate duality gap in the interior-point method
reached $0.01$.
(Although this is a similar stopping criterion to the one used 
for our method,
it is not equivalent to, nor a sufficient condition for,
having a relative gap less than $0.02$.)
\begin{table}
	\begin{center}
    \begin{tabular}{|c|c|}
    \hline 
    method              & solve time (s) \\ \hline\hline
    flow update         & $1.2$          \\ \hline
    distributed method  & $6268$         \\ \hline
    full CR problem     & $45953$        \\ \hline
  \end{tabular}
 \end{center}
 \caption{
    The timing results for the numerical example.
    }
 \label{optV-runtime}
\end{table}

For this example, our distributed method is about $7 \times$ faster 
than Mosek.  This timing advantage will of course scale with 
the number of cores or threads, since each iteration of our method is
trivially parallelizable, whereas the sparse matrix factorization that
dominates each iteration of an interior-point method is not.
By scaling the problem size just a bit larger, we reach a point where
direct solution of the CR is problem is no longer possible;
the distributed method, however, will easily handle (much) 
larger problems.

\section{Extensions}

\paragraph{Capacitated nodes.}
In our formulation,
only the links have capacity limits.
However, in many practical cases,
nodes may also have capacity limits,
with a corresponding reservation cost.

Given a problem with a capacitated node,
we show how to formulate an equivalent problem,
with only capacitated edges.
Consider a capacitated node $i$ with capacity $a$.
We can define new graph,
with node $i$ replaced by two nodes, $i_1$ and $i_2$;
all edges previously entering $i$ now enter $i_1$,
and all edges previously leaving $i$ now leave $i_2$.
We then connect $i_1$ to $i_2$ with a single new edge,
with capacity $a$ and edge reservation price $0$.

\paragraph{General reservation costs.}
The cost of reserving capacity on edge $j$
can be any convex function $\phi_j$
of the reservation $r_j$.

%

\paragraph{Multicommodity network.}
We can extend (\ref{e-cr-lp}) to handle $T$ different types of commodities.
The multicommodity capacity reservation problem is
\begin{equation}
\begin{array}{lll}
\mbox{minimize} & p^T r \\
\mbox{subject to} 
  & Af^{(k,t)} = s^{(k,t)}, \quad 0 \leq f^{(k,t)} 
      & \quad k = 1,\dots,K, \quad t=1,\dots,T \\
  & \sum_{t=1}^T f^{(k,t)} \leq r, & \quad k = 1,\dots,K, \\
  & r \leq c.
\end{array}
\label{e-cr-multi}
\end{equation}
The variables are the reservations $r\in\reals^m$
and $f^{(k,t)}\in \reals^m$
for $k=1,\dots,K$ and $t=1,\dots,T$.
The vector is $s^{(k,t)}$
is the source vector associated with type $t$ under scenario $k$.
We note that the scenario prices of \S\ref{s-lower-bounds}
extend immediately to the multicommodity case.

\newpage

\bibliographystyle{alpha}
\bibliography{opt_cap_res}

\newpage

\appendix

\section{Solution of the reservation update subproblem}
\label{a-res-update}
Here we give a simple method to carry out the
reservation update problem (\ref{e-res-update}).
This section is mostly taken from 
\cite[\S6.4.1]{parikh2014proximal}.

We can express (\ref{e-res-update}) as
\BEQ
\begin{array}{ll}
\mbox{minimize} & \beta t + (1/2)\|x-u\|^2 \\
\mbox{subject to} 
                 & x_i \leq t, \quad i = 1,\ldots K,
\end{array}
\EEQ
Here there variables are $x\in\reals^K$ and $t\in\reals$.
(The variable $x$ corresponds to $f_j$ in (\ref{e-res-update}),
and the parameters $u$ and $\beta$ 
correspond to $\alpha f_j + (1-\alpha \tilde f_j)+\pi_j/\rho$ and $p_j/\rho$, 
respectively.)

The optimality conditions are
\[
x^\star _i \leq t^\star,
\quad
\mu^\star_i \geq 0,
\quad
\mu_i^\star (x^\star_i - t^\star) = 0,
\quad
(x^\star_i - u_i) + \mu^\star_i = 0,
\quad
\ones^T \mu^\star = \beta,
\]
where $x^\star$ and $t^\star$ are the optimal primal variables,
and $\mu^\star$ is the optimal dual variable.
If $x_i < t$,
the third condition implies that $\mu^\star_i = 0$,
and if $x_i^\star = t^\star$,
the fourth condition implies that 
$\mu^\star_i = u_i - t^\star$.
Because $\mu_i^\star \geq 0$,
this implies
$\mu^\star_i = (u_i - t^\star)_+$.
Substituting for $\mu_i^\star$ in the fifth condition gives
\[
\sum_{i=1}^n (u_i - t^\star )_+ = \beta,
\]
We can solve this equation for $t^\star$
by first finding an index $k$ such that
\[
u_{[k+1]} - u_{[k]} \leq  ( u_{[k]} - \beta) / k
\leq u_{[k]} - u_{[k-1]}
\]
where $u_{[i]}$ is the sum of the $i$ largest elements of $u$.
This can be done efficiently by first sorting the elements of $u$,
then computing the cumulative sums in the descending order
until the left inequality in the above holds.
Note that in this way there is no need to check the second inequality,
as it will always hold.
(The number of computational operations required to sort $u$
is $K\log K$,
making this the most computationally expensive step of the solution.)
With this index $k$ computed,
we have
\[
t^\star = ( u_{[k]} - \beta) / k.
\]
We then recover $x^\star $
as $x^\star_i = \min\{t^\star, u_i\}.$

\section{Iterate properties}
\label{a-iterate-prop}
Here we prove that, for any $l=1,2,\dots$,
if $\pi^{(k)}$ are the columns of $\Pi(l)$,
optimality condition \ref{c-price-decomp} is satisfied,
\ie, $\Pi(l)\ones = p$ and $\Pi(l) \geq 0$.
Additionally, if $f^{(k)}$ are taken to be the columns of $F(l)$,
then condition \ref{c-zero-gap} is also satisfied,
\ie,
\[
\sum_{j=1}^m \sum_{k=1}^K \Pi(l)_{jk} \tilde F(l)_{jk}
=
p^T \max \tilde F(l).
\]

To see this, we first note that,
by defining $h:\reals^{m\times K}\to \reals$
such that $h(F) = p^T \max F$,
then the subdifferential of $h$ at $F$ is
\[
\partial h(F) =
\{ \Pi \in\reals^{m\times K} \mid
\Pi\ones = p, \; \Pi\geq 0, \; 
\Pi_{jk} > 0 \implies (\max F)_j = F_{jk}
\},
\]
\ie,
it is the set of matrices whose columns are scenario prices
and whose nonzero elements coincide with the elements of $F$
that are equal to the row-wise maximum value.

Now, from the reservation update equation,
we have
\[
\tilde F(l+1) = \argmin_{\tilde F} 
p^T \max (\tilde F) 
+ (\rho/2) \|\alpha F(l+1) + (1-\alpha)\tilde F(l)  + (1/\rho)\Pi(l) - \tilde F\|_F^2.
\]
The optimality conditions are
\[
\Pi(l) + \rho \big( - \tilde F(l+1) 
+ \alpha F(l+1) + (1-\alpha) \tilde F(l) \big)
\in \partial h\big(\tilde F(l+1) \big).
\]
Using the price update equation, this is
\[
\Pi(l+1) \in \partial h\big(\tilde F(l+1)\big).
\]

This shows that the columns of $\Pi(l+1)$ are indeed valid scenario prices.
Additionally,
we have (dropping the indices $(l+1)$ for $\Pi(l+1)$ and $\tilde F(l+1)$)
\[
\sum_{j=1}^m \sum_{k=1}^K \Pi_{jk} \tilde F_{jk}
=
\sum_{j=1}^m  (\max \tilde F)_j \sum_{k=1}^K \Pi_{jk}
=
\sum_{j=1}^m  (\max \tilde F)_j p_j
=
p^T \max \tilde F 
\]
The first step follows from the fact that the only
terms for which $\Pi(l+1)$ are positive,
and thus the only terms that contribute to the sum
are those for which the corresponding element in $\tilde F(l+1)$
is equal to the maximum in that row.
The second step follows from the fact that $\Pi(l+1)\ones = p$.

%
%
%
%


\end{document}